\documentclass[11pt,a4paper]{article}
\usepackage{jheppub_kim}

\usepackage{pdflscape}
\usepackage{amsmath}
\usepackage{amssymb}
\usepackage{dcolumn}
\usepackage{bm}
\usepackage{color}
\usepackage{epsfig}
\usepackage{amsfonts}
\usepackage{graphicx}
\usepackage{subfigure}
\usepackage{dcolumn}

\newcommand{\be}{\begin{equation}}
\newcommand{\ee}{\end{equation}}
\newcommand{\bea}{\begin{eqnarray}}
\newcommand{\eea}{\end{eqnarray}}

\def\be{\begin{equation}}
\def\ee{\end{equation}}
\def\bea{\begin{eqnarray}}
\def\eea{\end{eqnarray}}

\begin{document}

%%%%%%%%%%%%%%%%%%%%%%%%%%%%%%%%%%%%%%%%%%%%%%%%%%%%%%%%
\title{Cosmology with higher-derivative matter fields}
%%%%%%%%%%%%%%%%%%%%%%%%%%%%%%%%%%%%%%%%%%%%%%%%%%%%%%%%

\author[a,b]{Tiberiu Harko}
\author[c]{Francisco S. N. Lobo}
\author[d,e]{Emmanuel N. Saridakis}

\affiliation[a]{Department of Physics, Babes-Bolyai University, Kogalniceanu Street, 
Cluj-Napoca 400084, Romania}

\affiliation[b]{Department of Mathematics, University College London, Gower
Street, London
WC1E 6BT, United Kingdom}

\affiliation[c]{Instituto de Astrof\'{\i}sica e Ci\^{e}ncias do Espa\c{c}o, Faculdade de
Ci\^encias da Universidade de Lisboa, Edif\'{\i}cio C8, Campo Grande,
P-1749-016 Lisbon, Portugal}

\affiliation[d]{Physics Division, National Technical University of Athens,
15780 Zografou Campus,  Athens, Greece}

\affiliation[e]{Instituto de F\'{\i}sica, Pontificia Universidad  Cat\'olica
de Valpara\'{\i}so, Casilla 4950, Valpara\'{\i}so, Chile}

\emailAdd{t.harko@ucl.ac.uk}
\emailAdd{fslobo@fc.ul.pt}
\emailAdd{Emmanuel$_-$Saridakis@baylor.edu}

%%%%%%%%%%%%%%%%%%%%%%%%%%%%%%%%%%%%%%%%%%%%%%%%%%%%%%%%
\abstract{
 We investigate the cosmological implications of a new class of modified
gravity, where the field equations generically include higher-order
derivatives of the matter fields, arising from the introduction of
non-dynamical auxiliary fields in the action. Imposing a flat, homogeneous
and isotropic geometry we extract the Friedmann equations, obtaining an
effective dark-energy sector containing higher derivatives of the matter
energy density and pressure. For the cases of dust, radiation, and stiff
matter we analyze the cosmological behavior, finding accelerating, de
Sitter, and non-accelerating phases, dominated by matter or dark energy.
Additionally, the effective dark-energy equation-of-state parameter can be
quintessence-like, cosmological-constant-like, or even phantom-like.
The detailed study of these scenarios may provide signatures that could
distinguish them from other candidates of modified gravity.}
%%%%%%%%%%%%%%%%%%%%%%%%%%%%%%%%%%%%%%%%%%%%%%%%%%%%%%%%


\maketitle

\section{Introduction}

\label{Introduction}

Since the discovery of the late-time accelerated expansion of the Universe,
there has been much effort into understanding the perplexing nature of the
cosmic acceleration and of gravity itself. In a first direction, one can
introduce the dark-energy concept, in the context of scalar fields, as these
are popular building blocks used to construct models of present-day
cosmological acceleration. They are appealing because such fields are
ubiquitous in theories of high energy physics beyond the standard model and,
in particular, are present in theories which include extra spatial
dimensions, such as those derived from string theories. Thus, the
modification of the Universe content \cite{Copeland:2006wr} is materialized,
in general, by the addition of
extra dynamical scalar fields, for instance canonical (quintessence)
\cite{Ratra:1987rm,Wetterich:1987fm,Liddle:1998xm,
Guo:2006ab,Dutta:2009yb,Harko:2013gha},
phantom \cite{Caldwell:1999ew,Caldwell:2003vq,Nojiri:2003vn,Onemli:2004mb,
 Saridakis:2008fy}, their combination
\cite{Cai:2009zp,Guo:2004fq,Setare:2008pz},
K-essence \cite{ArmendarizPicon:2000ah,Chimento:2003ta}, Galileon
\cite{Nicolis:2008in}, etc.
On the other hand, one can modify the gravitational sector
\cite{Capozziello:2011et}, for instance constructing
$f(R)$ gravity
\cite{Capozziello:2002rd,Chiba:2003ir,Allemandi:2005qs,Nojiri:2006gh,
Nojiri:2007as,
Amendola:2006we,Starobinsky:2007hu,LanahanTremblay:2007sg,Boehmer:2007kx,
Harko:2011nh}, $f(T)$
gravity
\cite{ Bengochea:2008gz,Linder:2010py,Chen:2010va,Iorio:2012cm},
Weyl-Cartan-Weitzenb\"{o}ck gravity \cite{W1,W2}, Gauss-Bonnet gravity
\cite{Nojiri:2005vv,Koivisto:2006xf,DeLaurentis:2014oja}, Ho\v{r}ava-Lifshitz gravity
\cite{Horava:2009uw,Kiritsis:2009sh,Saridakis:2009bv,Bogdanos:2009uj,Saridakis:2012ui},
nonlinear massive gravity
\cite{deRham:2010ik,Hinterbichler:2011tt,Cai:2012ag,deRham:2014zqa}, etc,
which apart from the cosmological motivation has the additional advantage of
an
expected improved ultra-violet and quantum behavior of the theory
\cite{Stelle:1976gc}.
Furthermore, one could also construct combinations of these two directions,
such as in scalar-tensor theories
\cite{Uzan:1999ch,Amendola:1999qq,Fujii:2003pa}, in generalized Galileons
\cite{DeFelice:2010nf,Deffayet:2011gz,DeFelice:2011bh,Leon:2012mt}, etc. The
common feature of all the above theories is that they incorporate additional
degrees of freedom relatively to General Relativity and the standard
model.

Another interesting approach is that one could handle the gravitational and
matter sectors ``democratically'', that is modifying the matter part in the
Lagrangian too, along with its coupling to gravity
\cite{Bertolami:2007gv,Bertolami:2008ab,Bertolami:2008zh,Bertolami:2009ic,
Harko:2008qz,Harko:2010mv,Harko:2012hm,Wang:2012rw,Harko:2011kv,
Haghani:2013oma, Odintsov:2013iba, Harko:2014sja, Harko:2014aja}.
Generally, these theories lead to non-geodesic motion, which takes place in
the
presence of an extra force orthogonal to the four-velocity. The Newtonian
limit of the equation of motion was also considered, and a procedure for
obtaining the energy-momentum tensor of the matter was presented. On the
other hand, the gravitational field equations are equivalent to the Einstein
equations of the $f(R)$ model in empty spacetime, but differ from them, as
well as from standard General Relativity, in the presence of matter.
Therefore, the predictions of these models could lead to some major
differences,
as compared to the predictions of standard General Relativity, or its
extensions ignoring the role of matter, in several problems of current
interest, such as cosmology, gravitational collapse or the generation of
gravitational waves. The study of these phenomena may also provide some
specific signatures and effects, which could distinguish and discriminate
between the various theories of modified gravity.

Having these in mind, one could try to construct theories with additional
auxiliary fields which can alter the dynamics, however being themselves
non-dynamical and thus without altering the degrees of freedom
\cite{Pani:2013qfa,Banados:2013vya,Guo:2014bxa} (for similar constructions see  
\cite{Capozziello:2010uv,Cai:2010zma,Cai:2011bs}). Interestingly enough, the
requirements of satisfying the
weak equivalence principle and allowing for a covariant Lagrangian
formulation, lead the field equations of these theories to include, in
general, higher-order derivatives of the matter fields. Although this feature
places tight observational constraints on these theories, it is clear
that they correspond to novel modified classes.

The plan of the work is the following: In Section \ref{model}, we present
the gravitational modification with non-dynamical auxiliary fields,
extracting the corresponding cosmological equations. In Section
\ref{Solutions}, we solve the cosmological field equations and we provide the
cosmological
behavior for various matter equations of state. Finally, in  Section
\ref{Conclusions} we briefly discuss and conclude our paper.

\section{Cosmology with higher-derivative matter fields }
\label{model}

In this section we extract the cosmological equations of the scenario at
hand. In a first subsection, we briefly present the underlying gravitational
theory, namely gravity with auxiliary fields. Then, we apply it in a
cosmological framework, providing the Friedmann equations and defining the
basic observables.

\subsection{Gravity with auxiliary fields}
\label{model2.1}

The key point of the scenario at hand is that one introduces auxiliary
fields whose equations of motion are not dynamical \cite{Pani:2013qfa}. Thus,
one can use these equations to eliminate the auxiliary fields. In this way,
the equations of motion for the usual matter fields are not modified (which
is a significant advantage when it comes to the observational tests of the
theory), however the gravitational equations of motion do change, and in
particular they acquire higher order derivatives of the matter fields.
More specifically, the field equations can be written, without loss of
generality, as $G_{ab}+\Lambda g_{ab}=T_{ab}+S_{ab}({\bf g},{\bf T})$, where
the second rank tensor $S_{ab}({\bf g},{\bf T})$ depends explicitly on the
metric ${\bf g}$ and the matter fields (${\bf T}$ in general). The tensor
$S_{ab}$ is imposed to vanish in vacuum and be divergence-free, in virtue of
the Bianchi identity and the conservation of the matter fields, namely
$\nabla_a T^{ab}=0$.
Thus, in the present context, these theories with auxiliary fields consist in
modifying the Einstein field equation through the addition of a
divergence-free tensor, that vanishes in the vacuum, and depends on the
metric, the matter energy-momentum tensor, and its derivatives.
 
The above conditions place strict requirements on the second-rank tensor under
consideration. More specifically, the tensor $S_{ab}$, up to fourth order in the 
derivatives, is 
given by (we refer the reader to Ref. \cite{Pani:2013qfa} for more details)
\begin{eqnarray}
S_{ab} &=& \alpha_1\, g_{ab}\,T  
 + \alpha_2\, g_{ab}\, T^2 + \alpha_3 \, T\, T_{ab} + \alpha_4\, g_{ab}\, T_{cd}\,T^{cd}  
 + \alpha_5 \, T^c\,_a\,T_{cb}  
    \nonumber  \\
 &&+ \beta_1\, \nabla_a\nabla_b\,T + \beta_2 \, g_{ab}\, \Box\,T   +  \beta_3\, 
\Box\,T_{ab} + 2\beta_4\, \nabla^c\nabla_{(a}\,T_{b)c}+\ldots\,.
 \label{eqexp}  
\end{eqnarray}
Thus, in order to obtain $\nabla_a  S^{ab}=0$, to the considered order of expansion, one 
needs to 
impose the conditions 
\begin{eqnarray}
\alpha_1 = -\beta_1\,  \Lambda,  \quad  4\alpha_2 = 
\left(1+2\alpha_1\right)(\beta_1-\beta_4), \quad  \alpha_3 = 
\beta_4\left(1+2\alpha_1\right)-\beta_1, \nonumber \\ 2\alpha_4 = \,\beta_4\,, \quad  
\alpha_5 = - 2\beta_4\, ,  \quad  \beta_2 = - \beta_1,  \quad  \beta_3 = - \beta_4. 
\label{bonds}
\end{eqnarray} 
We refer the reader to Appendix (\ref{app}) for specific details.

Hence, independently of the specific initial action, the resulting Einstein equations 
become  \cite{Pani:2013qfa}
\begin{eqnarray}
G_{ab} &=& T_{ab} - \Lambda g_{ab}      -
\beta_1 \Lambda\, g_{ab}\,T + \frac{1}{4}\left(1-2\beta_1
\Lambda\right)(\beta_1-\beta_4)\, g_{ab}\, T^2  \notag \\
&&
+ \left[\beta_4\left(1-2\beta_1\, \Lambda\right) - \beta_1\right] \, T\,
T_{ab} + \frac{1}{2}\,\beta_4 \, g_{ab}\, T_{cd}\, T^{cd}  - 2\beta_4\,
T^c\,_a\,T_{cb} \notag \\
&& + \beta_1\, \nabla_a\nabla_b\,T - \beta_1\,
g_{ab}\, \Box\,T  - \beta_4 \, \Box\,T_{ab} + 2\beta_4\,
\nabla^c\nabla_{(a}\,T_{b)c}
+\ldots\,,  \label{fieldeq}
\end{eqnarray}
where $g_{ab}$ is the spacetime metric, $\Lambda$ the cosmological constant,
$T_{ab}$ the matter energy-momentum tensor, $T$ its trace, and we have set
Newton's constant to $1/8\pi $ (we refer the reader to \cite{Pani:2013qfa}
for more details). The divergence of the extra terms is zero, and the coefficients now 
only depend 
on the $\beta_1$ and $\beta_4$ model 
parameters, and it is
clear that when $\beta_1=\beta_4=0$ we recover the standard Einstein
equations. In the above equation one can clearly see that the
introduction of auxiliary fields indeed leads to gravitational equations
which include higher derivatives of the energy momentum tensor and its
trace.

Before proceeding to the cosmological investigation of the above scenario, let us make a 
comment on the subtle issue of considering the trace of the energy-momentum tensor, $
T$, in the Lagrangian. Indeed, note that the most obvious generalization of Newtonian 
gravity, based on Poisson's equation $\nabla^2 \phi=4\pi G \rho$ should be $\Box 
\phi=4\pi 
G {T}$, where the trace of the energy-momentum tensor acts as a source for the scalar 
field. This has several implications, such as: (i) the theory is nonlinear as the scalar 
field possesses a non-zero trace for the energy-momentum tensor, which appears in the 
right-hand-side of the above relationship; (ii) the scalar field also couples to the 
cosmological constant, as the latter has a non-zero energy-momentum trace, thus rendering 
a dynamical effective cosmological constant. In this context, the subtle issue is that 
the trace of the energy-momentum tensor should not be written a priori inside the 
Lagrangian but should be derived by varying the Lagrangian. Thus, the correct coupling 
would be the one with the properly defined trace, leading then to an infinite loop of
variations \cite{RNpaper,RNpaper1,Tamanini:2013aca,Ayuso:2014jda}. To this effect, 
consider a system consisting of gravitational 
fields $g_{\mu\nu}$, radiation fields, and a scalar field $\phi$ which couples to the 
trace of the energy-momentum tensor of all the fields, including its own 
\cite{RNpaper,RNpaper1,Tamanini:2013aca,Ayuso:2014jda,Padmanabhan:2002ji,Sami:2002se}. 
Furthermore, consider the 
zeroth order action given by
\begin{equation}
A^{(0)}=A_{\rm grav}+A^{(0)}_{\phi}+A^{(0)}_{\rm int}+A_{\rm rad}.
  \label{zeroaction}
\end{equation}
The respective terms are defined in the following manner
\begin{eqnarray}
A_{\rm grav}&=&\frac{1}{16\pi G}\int d^4x \sqrt{-g}\;R + \int d^4x \sqrt{-g}\;\Lambda \,,
   \\
A^{(0)} &=& \frac{1}{2} \int d^4x \sqrt{-g}\; \nabla^\mu \phi \, \nabla_\mu \phi   \,,
   \\
A^{(0)}_{\rm int}&=& \int d^4x \sqrt{-g}\;f(\phi/\phi_0)\,T \,,
\end{eqnarray}
where the cosmological constant is included. $\eta$ determines the strength of the 
interaction between the scalar field and the trace of the energy-momentum tensor, the 
constant $\phi_0$ is introduced for dimensional convenience, and finally note that the 
radiation term is traceless. Now, in order to take into account the back-reaction of the 
scalar field on itself, the trace of the scalar field, i.e. ${T}_{\phi}=-\nabla^\mu \phi 
\, \nabla_\mu \phi$, must be added to the total trace $T$. Next, the 
addition of ${T}_{\phi}$ in the interaction term $A^{(0)}_{\rm int}$ further modifies the 
energy-momentum tensor ${T}^{\mu\nu}_{\phi}$, and consequently modifies the trace 
${T}_{\phi}$. In conclusion, an infinite iteration needs to be performed to arrive at the 
correct action, where the complete action is obtained by summing up all the terms 
\cite{RNpaper,RNpaper1,Tamanini:2013aca,Ayuso:2014jda}.

Nevertheless, one should emphasize that the full action can be deduced by a simple 
consistency argument \cite{Padmanabhan:2002ji,Sami:2002se}. As the above-mentioned 
iteration is to modify the expressions for $A_{\phi}$ and $A_{\Lambda}$, one could 
consider the following as an ansatz for the full action
\begin{eqnarray}
A=\frac{1}{16\pi G}\int d^4x \sqrt{-g}\;R + \int d^4x \sqrt{-g}\;\alpha(\phi) \Lambda
    \nonumber \\
+\frac{1}{2} \int d^4x \sqrt{-g}\; \beta(\phi) \nabla^\mu \phi \, \nabla_\mu \phi   +
A_{\rm rad} \,,
 \label{itaction}
\end{eqnarray}
where the functions $\alpha(\phi)$ and $\beta(\phi)$ represent the effect of the 
iteration of the interaction term. Note that taking into account the action 
(\ref{itaction}), the energy-momentum tensor for the scalar field $\phi$ and the 
cosmological constant $\Lambda$ is now given by 
${T}^{\mu\nu}=\alpha(\phi)\Lambda g^{\mu\nu} + \beta(\phi) (\nabla^{\mu} \phi 
\nabla^{\nu} 
\phi -\frac{1}{2}g^{\mu\nu}  \nabla^{\alpha} \phi \nabla_{\alpha} \phi  )$, which 
provides the following trace: ${T}_{\rm tot}=  4 \alpha (\phi ) \Lambda - \beta(\phi) 
\nabla^{\alpha} \phi \; \nabla_{\alpha} \phi $. 
Using the latter trace equation, and equating the actions (\ref{zeroaction}) and 
(\ref{itaction}), 
the functions $\alpha(\phi)$ and $\beta(\phi)$ are given by
\begin{eqnarray}
\alpha(\phi)=(1+4\eta f)^{-1}\,, \qquad  \beta(\phi)=(1+2\eta f)^{-1}\,,
\end{eqnarray}
respectively. The cosmological behaviour was further extensively explored 
\cite{Padmanabhan:2002ji,
Sami:2002se}. Now, this model is conceptually attractive as it correctly accounts for the 
coupling of the scalar field and the trace of the energy-momentum tensor. However, there 
are several problems \cite{Padmanabhan:2002ji}, namely: (i) one does not arrive at a 
viable model using natural initial conditions without fine-tuning the parameters; (ii) 
due 
to the coupling of the scalar field to the trace of all the sources, it also couples and 
consequently  kills off the dust-like matter, rendering the present universe 
radiation-dominated; (iii) the latter point reduces the age of the universe and creates 
difficulties for the formation of structure. It has been emphasized that these 
problems may be solved by invoking a suitable potential $V(\phi)$, but this eliminates 
the naturalness of the model \cite{Padmanabhan:2002ji,Sami:2002se}.

In the present context, one may also argue that the theory analyzed in this work suffers 
form similar drawbacks in that one should not place the trace of the energy-momentum 
tensor inside the Lagrangian, but he should derive the latter by variation. Thus, the 
correct coupling would be the one with the properly defined trace, leading then to an 
infinite loop of variations. However, we emphasize that we may consider the theory at a 
phenomenological level, where we have postulated the trace in the Lagrangian, and 
extensively explore its consequences and cosmological behavior. This is what we do in the 
rest of the manuscript.

\subsection{Cosmological equations}
\label{Cosmology}

In order to apply the above modified gravitational theory in cosmology we
consider the usual homogeneous and isotropic geometry, given by the flat
Friedmann-Robertson-Walker (FRW) metric
\begin{equation}
ds^2= - dt^2 + a^2(t)\,\delta_{ij} dx^i dx^j,
\end{equation}
where $a(t)$ is the scale factor. Additionally, concerning the matter
energy-momentum tensor, we use the standard form of a perfect fluid, namely
$T_{ab}=(\rho+p) U_{a} U_{b}+p \, g_{ab}$, where $\rho$ and $p$ are
respectively the matter energy density and pressure.

Inserting these in the field equations (\ref{fieldeq}), we obtain the
modified Friedmann equations as
\begin{eqnarray}\label{eqH}
3H^2 &=&  (\rho +\Lambda)+3H \left[
\beta_1 ( \dot{\rho} - 3\dot{p}) - \beta_4 ( \dot{\rho} +
2\dot{p})  \right]   - \beta_4 \ddot{\rho}   \nonumber \\
&&
+ \frac{1}{4}\Big[  3 \beta_1 (1+
2\beta_4 \Lambda)
(\rho^2 -3 p^2)+3\beta_4 (\rho^2 + p^2) - 12 \beta_1 (\beta_1 + \beta_4)
\Lambda  \rho p
   \nonumber \\
&&
 - 6 (\beta_1 - \beta_4) \rho p
-4\beta_1 \Lambda  (\rho - 3p)  + 2\beta_1^2
\Lambda (\rho^2 + 9p^2)
\Big],
 \end{eqnarray}
 and
\begin{eqnarray}\label{eqdH}
\dot{H}\left[1+\beta
_4\left(\rho +p\right)\right]&=& - \frac{1}{2}(\rho + p)
- 3H^2  \beta_4 (\rho + p)
- \frac{H}{2} \left[   \beta_1 (\dot{\rho} - 3 \dot{p})  -  \beta_4
(\dot{\rho} + 3 \dot{p})  \right]
 \nonumber  \\
&&
-\frac{1}{2}\Big[  (3\beta_1 + \beta_4)
\ddot{p} -(\beta_1 + \beta_4) \ddot{\rho} + \beta_4 (\rho^2 + p^2)
    \nonumber  \\
&& + \beta_1
(1 + 2 \beta_4 \Lambda) (\rho^2 -3
p^2)  -2 (\beta_1 -\beta_4 +2
\beta_1 \beta_4 \Lambda) \rho p  \Big],
\end{eqnarray}
respectively. The above equations can be re-written in the form of the usual
Friedmann
equations as
\begin{equation}
\label{Fr11}
3H^{2}=\rho +\rho _{DE},
\end{equation}%
\begin{equation}
\label{Fr22}
-2\dot{H}=   \rho +\rho _{DE}+p+p_{DE}  ,
\end{equation}%
where we have introduced the effective dark-energy sector  with energy
density and pressure, $\rho_{DE}$ and $p_{DE}$, respectively, defined as
\bea
\rho _{DE}&=&\Lambda + \frac{1}{4}\Big[  3 \beta_1 (1+ 2\beta_4 \Lambda)
(\rho^2 -3 p^2)
+3\beta_4 (\rho^2 + p^2) - 12 \beta_1 (\beta_1 + \beta_4) \Lambda  \rho p - 6
(\beta_1 - \beta_4) \rho p
   \nonumber \\
&&-4\beta_1 \Lambda  (\rho - 3p)  + 2\beta_1^2 \Lambda (\rho^2 + 9p^2)
\Big]
+3H \left[ \beta_1 ( \dot{\rho} - 3\dot{p})  - \beta_4 ( \dot{\rho} +
2\dot{p})  \right]   - \beta_4 \ddot{\rho},
\label{rhoDE}
\eea
and
\begin{eqnarray}
p_{DE} &=&\frac{1}{4\left[ \beta
_{4}(p+\rho )+1\right] }
\Bigg\{-\beta
_{4}\rho ^{3}\left[ 2\beta _{1}\left( \beta _{1}+3\beta _{4}\right) \Lambda
+3\left( \beta _{1}+\beta _{4}\right) \right] -\left( \beta _{1}-3\beta
_{4}\right) \rho ^{2}\left( 2\beta _{1}\Lambda -1\right)
\nonumber\\
&& +4\rho \left( \beta
_{1}\Lambda -\beta _{4}\Lambda +6\beta _{4}H^{2}\right) +4 p
\left[   6\beta
_{4}H^{2}
- \left( 3\beta _{1}+\beta _{4}\right) \Lambda \right]
\nonumber\\
&& +
 \beta_{4} p\rho^2 \left[ \beta _{1}\left( 10\beta
_{1}\Lambda +6\beta _{4}\Lambda +3\right) -9\beta _{4}\right]  +
2 p\rho \left[ 6\beta
_{1}\left( \beta _{1}-\beta _{4}\right) \Lambda -\left( \beta _{1}+3\beta
_{4}\right) \right]
\nonumber\\
&& +3p^{2}\left[ \beta _{4}\rho
\left( -2\left( \beta _{1}-5\beta _{4}\right) \beta _{1}\Lambda +5\beta
_{1}-3\beta _{4}\right) -\left( \beta _{1}+\beta _{4}\right) \left( 6\beta
_{1}\Lambda +1\right) \right]
\nonumber\\
&&  -4\Lambda
+3\beta _{4}p^{3}\left[ 6\left( \beta _{4}-\beta _{1}\right) \beta
_{1}\Lambda +3\beta _{1}-\beta _{4}\right]
\nonumber\\
&&  + 4H\left[ \left( \beta _{1}-\beta
_{4}\right) \dot{\rho} (t)-3\left( \beta _{1}+\beta _{4}\right)
\dot{p}(t)\right] +4\left( 3\beta _{1}+\beta
_{4}\right)
\ddot{p}%
-4\left( \beta _{1}+\beta _{4}\right) \ddot{\rho}\Bigg\}.
\end{eqnarray}
Thus, we can calculate the equation-of-state parameter $w_{DE}$ of the
effective dark-energy sector  as
$w_{DE}\equiv p_{DE}/\rho _{DE}$, or using the Friedmann equations
(\ref{Fr11}) and (\ref{Fr22}) as
\be
w_{DE}=-\frac{2\dot{H} +p+3H^2}{3H^2-\rho }.
\label{wDE}
\ee
Moreover, one can see that given the matter energy conservation
\begin{equation}  \label{cons}
\dot{\rho}+3H\left( \rho +p\right) =0,
\end{equation}
the effective dark energy is also conserved, that is
\begin{equation}  \label{cons2}
\dot{\rho}_{DE}+3H\left( \rho_{DE} +p_{DE}\right) =0.
\end{equation}
Finally, we can define the deceleration parameter $q$,  namely
\begin{equation}
\label{qq}
q=-1-\frac{\dot{H}}{H^2}=\frac{1}{2}+\frac{3}{2}\frac{p+p_{DE}}{\rho +\rho
_{DE}}.
\end{equation}
The sign of $q$ indicates the decelerating/accelerating nature
of the cosmological expansion; cosmological models with
$q < 0$ are accelerating, while those having $q > 0$ experience
a decelerating evolution. Note that in terms of $q$ the dark-energy
equation-of-state parameter
reads
\be
w_{DE}=\frac{(2q-1)H^2-p}{3H^2-\rho},
\label{wDE2}
\ee
or, alternatively,
\be\label{wdei}
w_{DE}=\frac{1}{3}(2q-1)\left(\frac{\rho }{\rho _{DE}}+1\right)-\frac{p}{\rho
_{DE}}.
\ee

In order to be able to handle the cosmological equations of the scenario at
hand, namely the Friedmann equations (\ref{Fr11}) and (\ref{Fr22}) and the
conservation equations (\ref{cons}) (or (\ref{cons2})), we proceed as
follows. Firstly, from equation (\ref{cons}) we easily obtain
\begin{equation}
H=-\frac{\dot{\rho}}{3\left( \rho +p\right) },  \label{H1}
\end{equation}%
and
\begin{equation}
\dot{H}=\frac{\dot{\rho}\left( \dot{p}+\dot{\rho}\right) -(p+\rho
)\ddot{\rho%
}}{3(p+\rho )^{2}}.  \label{dH1}
\end{equation}%
 Then, substitution  into equations
(\ref{eqH}) and (\ref{eqdH}) provides the equations
\bea
\label{rho21}
\beta _4\ddot{\rho} &=&
\Lambda +\rho +\frac{\dot{\rho} \left[(3 \beta _1+2
\beta _4) \dot{p}+(\beta _4-\beta _1) \dot{\rho}\right]}{p+\rho}
 -\frac{\dot{\rho}^2}{3 (p+\rho )^2}
 \nonumber\\
&&  +\frac{1}{4} \Bigg\{-2 \rho  \left\{2
\beta _1 \Lambda +3 p [2 \beta _1
\Lambda  (\beta _1+\beta _4)+\beta _1-\beta _4]\right\} +\rho ^2 \left[2
\beta
_1
   \Lambda  (\beta _1+3 \beta _4)+3 (\beta _1+\beta
_4)\right]
    \nonumber\\
&& \ \ \ \ \ \ \ \ \ \ \  + 3p \left[4 \beta _1 \Lambda +p (6 \beta _1
\Lambda (\beta _1-\beta
_4)-3 \beta _1+\beta _4)\right]\Bigg\} ,
\eea
and
\bea
\label{rho22}
\ddot{\rho} &=& -\left(3\beta _{1}+5\beta
_{4}+\frac{2}{p+\rho }\right)^{-1}\Bigg\{
3p\left[ 2\rho (2\beta
_{1}\beta _{4}\Lambda +\beta _{1}-\beta _{4})-1\right] +3p^{2}\left[ \beta
_{1}(6\beta _{4}\Lambda +3)-\beta _{4}\right]
    \nonumber\\
&&
- \frac{3(3\beta _{1}+\beta _{4})(p+\rho
)^{2}\ddot{p}+\dot{p}\dot{\rho}\left[
(3\beta _{1}+5\beta _{4})(p+\rho )+2\right] -\dot{\rho}^{2}\left[ (\beta
_{1}-5\beta _{4})(p+\rho )-2\right] }{(p+\rho )^{2}}
\nonumber\\
&&\ \ \ \ \ \ \ \ \ \ \  -3\rho ^{2}(2\beta _{1}\beta
_{4}\Lambda +\beta _{1}+\beta _{4})-3\rho \Bigg\} ,
\eea
Finally, eliminating $\ddot{\rho}$ between equations (\ref{rho21}) and
(\ref{rho22}),
through equalizing the right hand sides, we obtain a differential equation
for $\rho$. Hence, if we additionally provide the matter equation of state,
assumed to be of barotropic form $p=p(\rho)$, this differential equation can
be solved to give $\rho(t)$. Lastly, $H(t)$ is then obtained from equation
(\ref{H1})
and then $\rho_{DE}(t)$ from equation (\ref{rhoDE}), $w_{DE}$ from equation
(\ref{wDE}) and
$q$ from equation (\ref{qq}).

\section{Cosmological solutions}
\label{Solutions}

In this section we explore the cosmological scenario with higher-matter
derivatives, for three particular cases of matter equations of state, namely
dust ($p=0$), radiation ($p=\rho /3$) and stiff fluid ($p=\rho$).

\subsection{Dust cosmological models}

As a first example we consider that the matter is in the form of dust,
namely with
$p=0$. In this case equations (\ref{rho21}) and (\ref{rho22}) are
respectively simplified to
\begin{equation}
\label{rho31}
\ddot{\rho }= \frac{3 \rho ^2 \left\{4 \Lambda +\rho  \left\{4-4 \beta _1
\Lambda +\rho  [2 \beta _1 \Lambda  (\beta _1+3 \beta _4)+3
   (\beta _1+\beta _4)]\right\}
   \right\}-4 \dot{\rho }^2 \left[3 (\beta
_1-\beta _4) \rho +1\right]}{12 \beta _4 \rho ^2},
\end{equation}
and
\bea\label{rho32}
\ddot{\rho }= \frac{3 \rho ^3 \left[\rho  \left(2 \beta _1 \beta _4 \Lambda
+\beta _1+\beta _4\right)+1\right]+\dot{\rho}^2 \left[2-(\beta _1-5
   \beta _4) \rho \right]}{\rho  \left[(3 \beta _1+5 \beta _4) \rho
+2\right]}.
   \eea
Hence, as we mentioned in the previous section, equality of the
right-hand-sides provides the differential 
equation for $\rho(t)$. If for convenience
we introduce the dimensionless variables
\begin{equation}
\tau=\frac{t}{2\beta _{4}^{1/2}},   \qquad  \theta =\beta _{4} \rho , \qquad 
k=\frac{\beta
_{1}}{\beta _{4}}, \qquad h= \beta _4^{1/2} H, \qquad \lambda = \beta
_{4}\Lambda ,
\end{equation}
the differential equation takes the following form
\begin{eqnarray}
\label{dens1}
\frac{d\theta }{d\tau }&=&-\theta
\left\{\theta
\left[3k(\theta +3\theta
k+3)+5\right]+2\right\}^{-1/2}
\Bigg\{3\theta \left(\theta +3\theta k+4\right)[3\theta
(k+1)+2]
    \nonumber\\
&& +6\lambda \Big\{ \theta \left\{
k\left[ \theta ^{2}(3k^2+14k+3)-4\theta (k+1)+2\right] +10\right\}
+4\Big\}
\Bigg\}^{1/2},
\end{eqnarray}%
where for physical reasons (monotonically decreasing energy density) we
have adopted the minus sign for the square root.
Additionally, calculating $H(t)$  from equation (\ref{H1}) and inserting
into equation (\ref{qq}) we extract the analytical expression for the
deceleration parameter as
\begin{eqnarray}
&&q = \Bigg\{ \left[6k\theta (\theta +3k\theta
+3)+4\right]\Big\{ \theta
\left[ \theta^{2}k^2(6k\lambda+28\lambda +9)+\theta^2 (6k\lambda
+12k+3)
\right.
       \nonumber\\
 &&
 \ \ \ \ \ \ \ \ \ \ \ \ \  \ \ \ \ \ \ \ \ \ \ \ \ \  \ \ \ \ \ \ \ \ \ \ \
\ \ \left. +2\theta (9k+7-4\lambda k^2-\lambda k )+
4\lambda(k+5) +8\right]
+8\lambda \Big\} \Bigg\} ^{-1}  \times
     \nonumber \\
&&\Bigg\{
3\theta^5(3k^2+k) \left[k^2(6k\lambda+28\lambda
+9)+6k(\lambda +2)+3\right]-\theta^2
\left[8k(89k+77)\lambda-112\right]
\nonumber \\
&&\ \ \  -\theta^3   \left[ 8k^2\lambda(21k+80) +72k^2
+2k(108\lambda -87)
-42\right]
\nonumber\\
&&\ \ \  +3\theta^4
\left[8k(15\lambda k^3   +   50\lambda k^2+11\lambda k)+8k(3k+1)\right]
+8\theta \lambda(5-44k) +16\theta
-32\lambda \Bigg\}.
\label{qgen}
\end{eqnarray}%
Note that for the specific case where the cosmological
constant is absent (namely
$\lambda =0$), that is when the effective dark energy sector constitutes
solely from the higher matter derivatives, we obtain
\begin{eqnarray}
\label{qm1}
q=\frac{1}{2}+\frac{3 (9 k \theta+4)}{6 k \theta (\theta+3 k \theta+3)+4}-
 \frac{6}{\theta+3 k \theta+4}-\frac{3}{3 (k+1) \theta +2}.
\end{eqnarray}
\begin{figure*}[tbp]
\centering
\includegraphics[width=7.5cm]{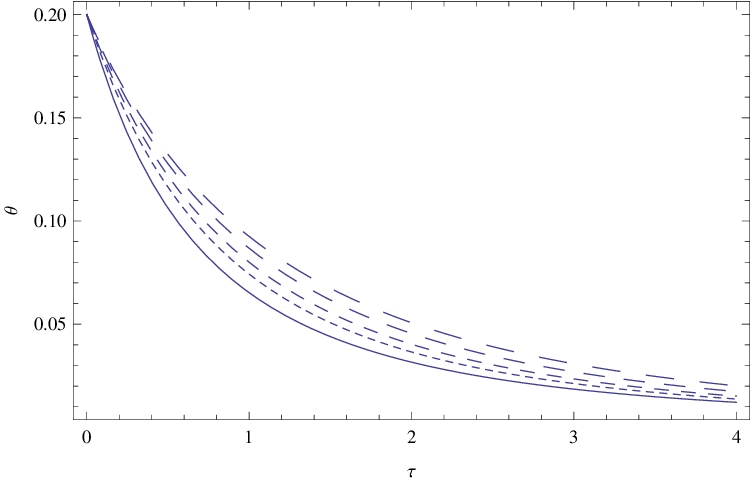}  %
\includegraphics[width=7.5cm]{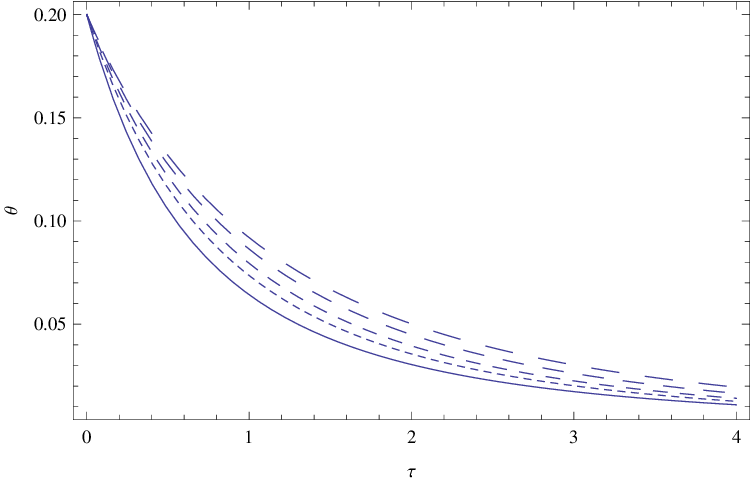}
\caption{{\it{Time evolution of the scaled matter energy density
$\protect\theta $ of the dust-matter
Universe, as a function of
$\protect\tau$,  for $\protect\lambda =0$ (left panel) and for
$%
\protect\lambda =0.004$ (right panel), for different values of $k$: $k=1$
(solid curve), $k=3$ (dotted curve), $k=5$ (short dashed curve), $k=9$
(dashed curve) and $k=16$ (long dashed curve). The initial value of the
energy density is $\protect\theta (0)=0.20$. }}}
\label{fig1}
\end{figure*}
\begin{figure*}[tbp]
\centering
\includegraphics[width=7.5cm]{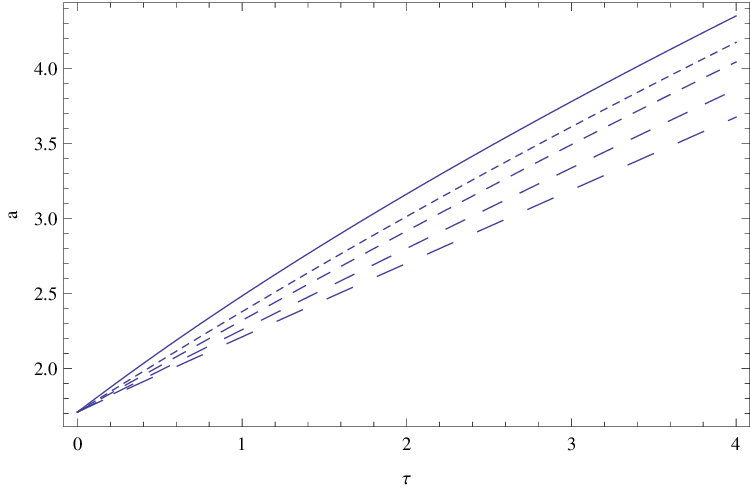}  %
\includegraphics[width=7.5cm]{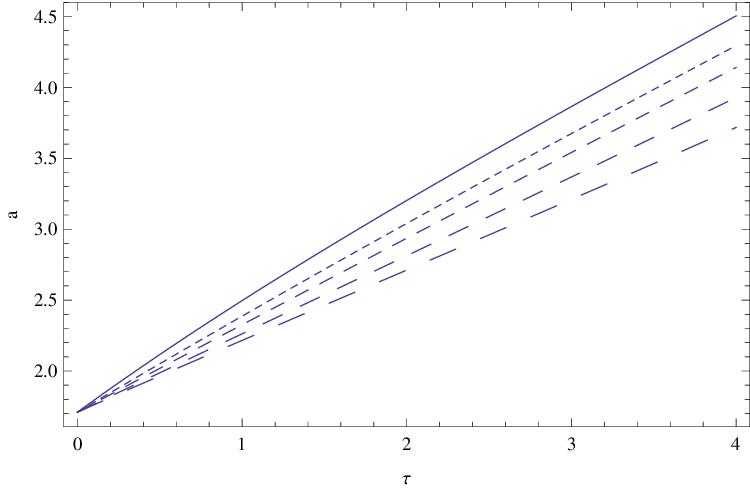}
\caption{{\it{Time evolution of the scale factor $a $ of the dust-matter
Universe, as a function of
$\protect\tau$, 
for $\protect\lambda =0$ (left panel) and for $\protect\lambda =0.004$
(right panel), for different values of $k$: $k=1$ (solid curve), $k=3$
(dotted curve), $k=5$ (short dashed curve), $k=9$ (dashed curve) and $k=16$
(long dashed curve). The initial value of the energy density is $\protect%
\theta (0)=0.20$. }}}
\label{fig2}
\end{figure*}
\begin{figure*}[tbp]
\centering
\includegraphics[width=7.5cm]{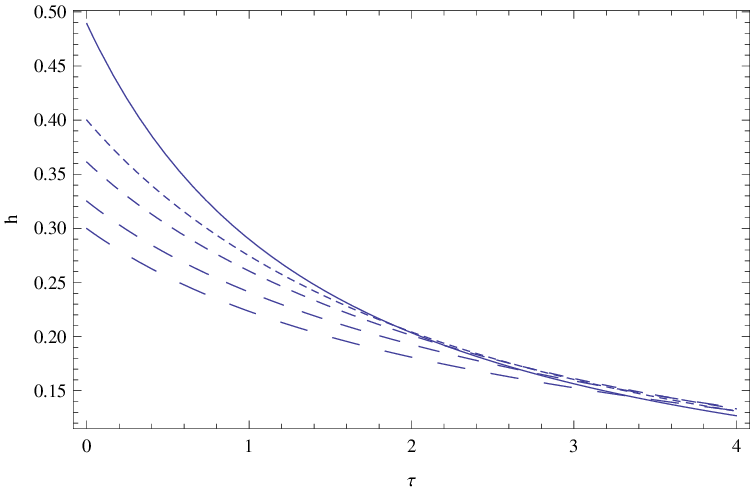}  %
\includegraphics[width=7.5cm]{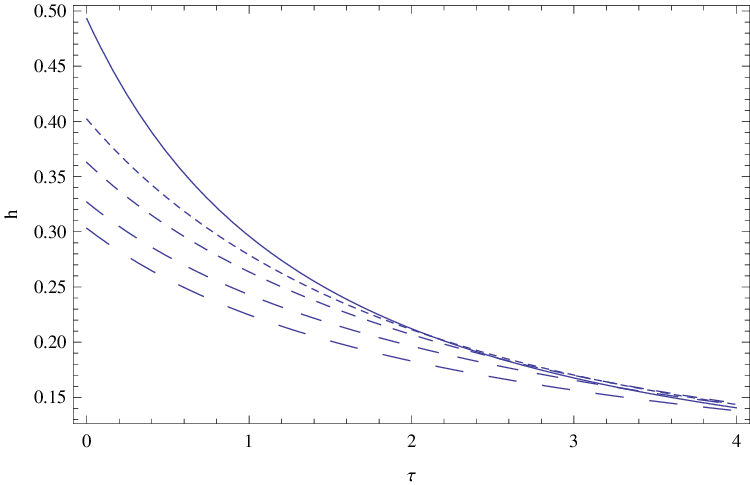}
\caption{{\it{Time evolution of the Hubble function $h $ of the dust-matter
Universe, as a function of
$\protect\tau$,  for $\protect\lambda =0$ (left panel) and for
$\protect\lambda =0.004$
(right panel), for different values of $k$: $k=1$ (solid curve), $k=3$
(dotted curve), $k=5$ (short dashed curve), $k=9$ (dashed curve) and $k=16$
(long dashed curve). The initial value of the energy density is $\protect%
\theta (0)=0.20$. }}}
\label{fig3}
\end{figure*}
\begin{figure*}[tbp]
\centering
\includegraphics[width=7.5cm]{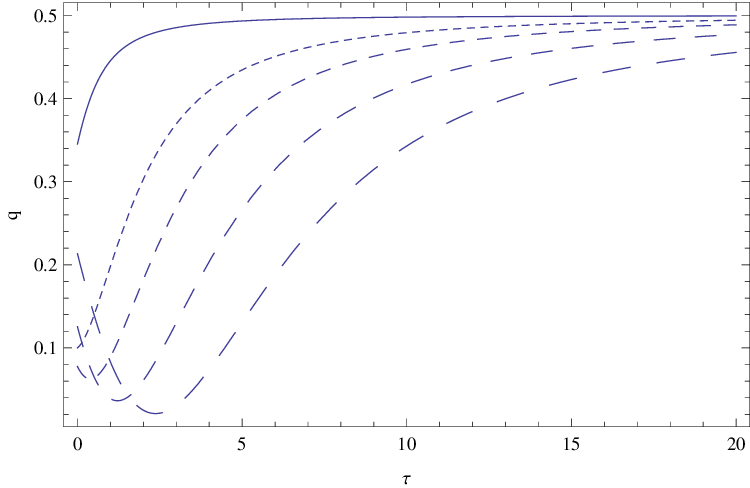}  %
\includegraphics[width=7.5cm]{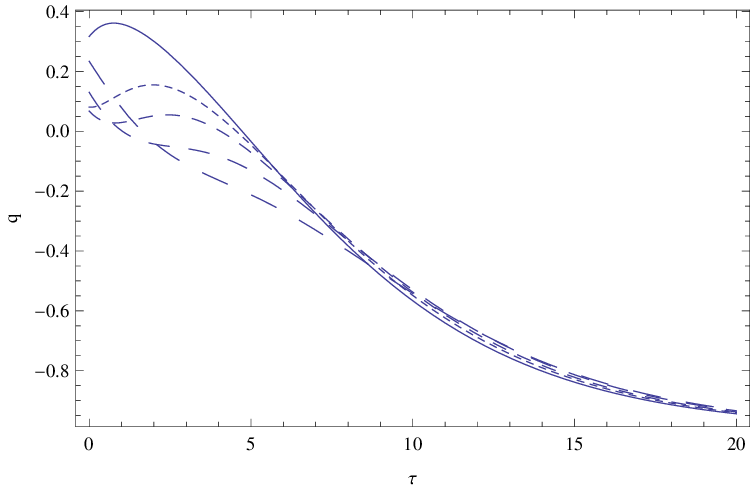}
\caption{{\it{Time evolution of the deceleration parameter $q$ of the
dust-matter
Universe, as a function of
$\protect\tau$,  for $\protect\lambda =0$ (left panel) and for $\protect%
\lambda =0.004$ (right panel), for different values of $k$: $k=1$ (solid
curve), $k=3$ (dotted curve), $k=5$ (short dashed curve), $k=9$ (dashed
curve) and $k=16$ (long dashed curve). The initial value of the energy
density is $\protect\theta (0)=0.20$. }}}
\label{fig4}
\end{figure*}
\begin{figure*}[tbp]
\centering
\includegraphics[width=7.5cm]{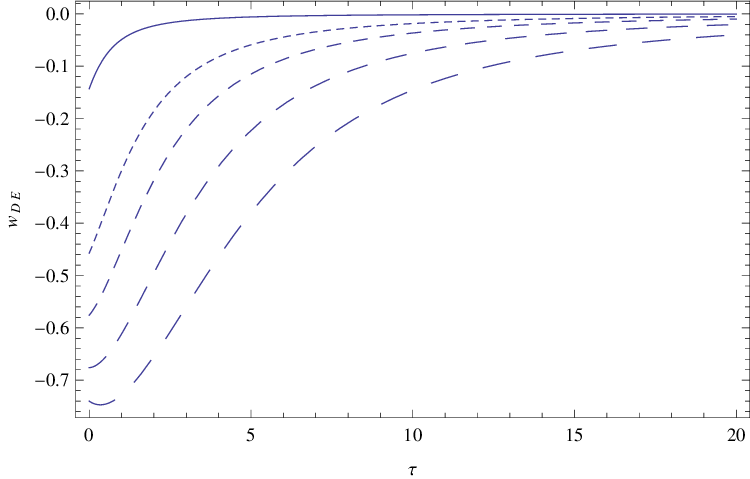}  %
\includegraphics[width=7.5cm]{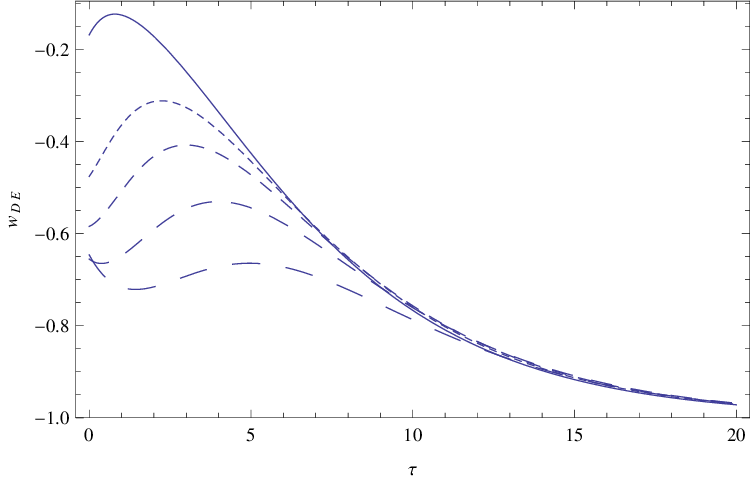}
\caption{{\it{Time evolution of the dark-energy equation-of-state parameter 
$w_{DE} $
of the dust-matter
Universe, as a function of
$\protect\tau$,  for $\protect\lambda =0$ (left panel) and for $\protect%
\lambda =0.004$ (right panel), for different values of $k$: $k=1$ (solid
curve), $k=3$ (dotted curve), $k=5$ (short dashed curve), $k=9$ (dashed
curve) and $k=16$ (long dashed curve). The initial value of the energy
density is $\protect\theta (0)=0.20$. }}}
\label{fig4a}
\end{figure*}

In order to present the above behavior more transparently, we
numerically solve equation (\ref{dens1}), and then we calculate the various
physical quantities as we described above. In Figs.~\ref{fig1}-\ref{fig4a}
we present the time evolution of the matter energy density, of the scale
factor, of the Hubble function, of the deceleration parameter and of the
dark-energy equation-of-state  parameter, for the cases $\lambda =0$ and
$\lambda =0.004$,  and for different
values of $k$.

First of all, as expected, the cosmological dynamics depends on the value of
$\lambda$. In particular, since in the scenario at hand the effective dark
energy sector is attributed to the higher derivatives of the matter fields,
when an explicit cosmological constant is absent ($\lambda=0$) we expect the
dark-energy equation-of-state parameter to become asymptotically zero at
late times, and the Universe to be non-accelerating. This behavior can be
clearly seen in the left graphs of Figs. \ref{fig4} and  \ref{fig4a}. On the
other hand, when $\lambda\neq0$, the explicit cosmological constant dominates
inside the effective dark energy sector, and in this case the
quintessence-like Universe at
late times transits to the accelerating phase, tending asymptotically to the
de Sitter evolution, with $w_{DE}\rightarrow-1$. This behavior can be
observed in the right graphs of Figs. \ref{fig4} and  \ref{fig4a}. 

The above asymptotic behaviors can be also analytically extracted from the
cosmological equations. In particular, setting $\lambda=0$ in  
(\ref{qm1}) leads immediately to $\lim_{\theta
\rightarrow 0}  q =1/2$. On the other hand,
for $\lambda > 0$, expression (\ref{qgen}) leads to  
$\lim_{\theta
\to 0}  q=-1$.

We close this subsection by mentioning that more complex behaviors of
$w_{DE}$ can also be achieved. In particular, according to (\ref{wdei}), if
during the cosmological expansion for some $\rho =\rho _{cr}$ 
%and $p=p_{cr}$
the condition
\be
\frac{1}{3}(2q-1)\left(\frac{\rho _{cr}}{\rho
_{DE}}+1\right)
%-\frac{p_{cr}}{\rho _{DE}}
=w_{DE}^{(cr)}=-1,
\ee
is satisfied, then at the corresponding time the phantom divide crossing
will be realized.

\subsection{The radiation dominated phase}

Let us now investigate the cosmological behavior of a radiation
dominated
Universe, with the matter pressure   satisfying the equation of state $
p=\rho /3$. The gravitational field equations (\ref{eqH}), (\ref{eqdH})
and (\ref{cons}) give
\begin{equation}  \label{rad1}
\ddot{\rho}= \frac{60 \beta _4 \rho \dot{\rho }^2+64 \beta _4 \rho ^4+48
\Lambda \rho ^2-9 \dot{\rho }^2+48 \rho ^3}{48 \beta _4 \rho ^2},
\end{equation}
\begin{equation}  \label{rad2}
\ddot{\rho }= \frac{30 \beta _4 \rho \dot{\rho }^2+32 \beta _4 \rho ^4+9
\dot{\rho }^2+24 \rho ^3}{3 \rho (8 \beta _4 \rho +3)},
\end{equation}
and
\begin{equation}
\dot{\rho }+4H\rho=0,
\end{equation}
respectively. It is interesting to note that for the radiation-dominated
Universe of this subsection the dynamics is determined solely by the
parameter $\beta_4$, with the parameter $\beta _1$ eliminated from the
equations. The consistency
condition requiring the equality of the right-hand-sides of equations
(\ref{rad1}) and (\ref{rad2}) provides the basic evolution equation
describing the
dynamical behavior, namely
\begin{equation}
\dot{\rho }=-4\rho \sqrt{ \frac{ 3 \Lambda +\left(3+8 \beta _4
\Lambda\right) \rho +4 \beta _4 \rho ^2 }{3 (4 \beta _4 \rho +3)}}.
\end{equation}
In the case $\Lambda =0$ and $\beta_4=0$, we obtain
\begin{equation}
\dot{\rho }=-\frac{4 \rho ^{3/2}}{\sqrt{3}},
\end{equation}
which yields the general solution
\begin{equation}
\rho =\frac{3\rho _0}{\left(2\sqrt{\rho _0}t+\sqrt{3}\right)^{2}},
\end{equation}
where we have used the initial condition $\rho (0)=\rho _0$. Therefore, in
this case we re-obtain the cosmological behavior of the radiation filled
Universes in standard cosmology, with 
\begin{eqnarray}
H(t)&=& \frac{\sqrt{\rho _0}}{2\sqrt{\rho
_0}t+\sqrt{3}}  , 
    \\
a(t)&=&a_0\sqrt{2t\sqrt{\frac{\rho_0}{3}}+1} ,
\end{eqnarray}
and $q=1$, respectively.

\subsection{Stiff fluid cosmology}

One of the most common equations of state of high energy density cosmological
matter, which has  been used extensively to study the properties of the
early Universe, is the linear barotropic equation of state, given by $p =
(\gamma - 1)\rho $, with $\gamma = \mathrm{constant} \in [1,2]$. A very
important subcase  is its so-called causal limit, corresponding to $\gamma =
2$, which gives the Zeldovich, or stiff-fluid equation of state $p = \rho $
\cite{60paper}.
The Zeldovich equation of state applies to densities significantly
higher than nuclear densities, $\rho > 10\rho _{\rm nuc}$, with $\rho _{\rm
nuc} =
10^{14}$ g/cm$^3$. From a field theoretical point of view the Zeldovich
equation of state can be obtained by constructing a relativistic
Lagrangian that allows bare nucleons to interact attractively via scalar
meson exchange and repulsively via the exchange of a more massive vector
meson \cite{60paper}. On the other hand, in the non-relativistic limit both the
quantum and
classical field theories yield Yukawa-type potentials. At the highest
densities
the vector-meson exchange dominates, and by using a mean field approximation
one can show that in the extreme limit of infinite densities the pressure
tends to the energy density, namely $p\rightarrow \rho $ \cite{60paper}. In this
limit the sound
speed $c_s^2 = dp/d\rho \rightarrow 1$, and hence this equation of state
satisfies the causality condition, with the speed of sound less than the
speed of light.

For a stiff fluid the field equations (\ref{eqH}), (\ref{eqdH}) and (%
\ref{cons})
provide respectively
\begin{equation}
\ddot{\rho }= \frac{12 \beta _1 \rho \dot{\rho }^2-36 \beta _1 \rho ^4+18
\beta _4 \rho \dot{\rho }^2+36 \beta _4 \rho ^4-\dot{\rho }^2+12 \rho ^3}{12
\beta _4 \rho ^2},
\label{eq11}
\end{equation}
\begin{equation}
\ddot{\rho }= \frac{-\beta _1 \rho \dot{\rho }^2+12 \beta _1 \rho ^4-5 \beta
_4 \rho \dot{\rho }^2-12 \beta _4 \rho ^4-\dot{\rho }^2-6 \rho ^3}{\rho (6
\beta _1 \rho -2 \beta _4 \rho -1)},
\label{eq22}
\end{equation}
and
\begin{equation}
\dot{\rho }+6H\rho=0.
\label{eq33}
\end{equation}
We mention that we have set to zero the explicit cosmological constant
(namely $\Lambda =0$), since it is negligible comparing to the high
densities that are needed to justify the Zeldovich stiff equation of state.

The basic evolution equation that describes the energy density  
evolution, obtained by equating the right hand sides of (\ref{eq11}) and
(\ref{eq22}), is given by
\begin{equation}
\label{finZ}
\dot{\rho }=-2 \sqrt{3} \sqrt{\frac{\rho ^3 \left[6 (\beta _1-\beta _4) (3
\beta _1+\beta _4) \rho ^2-(9 \beta _1+\beta _4) \rho +1\right]}{2 \rho
\left[-9
\beta _1-2 \beta _4+12 (\beta _1+\beta _4) (3 \beta _1+\beta _4) \rho
\right]+1}}.
\end{equation}
By introducing the set of dimensionless quantities $\left(\tau, \theta ,
k,h\right)$, defined as
\begin{equation}
\tau=\frac{2\sqrt{3}}{\beta _4^{1/2}}\;t, \qquad \theta = \beta _4 \rho,
\qquad k=
\frac{\beta _1}{\beta _4}, \qquad h=\frac{\beta _4^{1/2}}{2\sqrt{3}}H,
\end{equation}
equation (\ref{finZ}) takes the final form
\begin{equation}
\dot{\theta}=-\theta ^{3/2}\sqrt{\frac{\theta \left[6 (k-1) (3 k+1) \theta
-9 k-1\right]+1}{ 2 \theta \left[12 (k+1) (3 k+1) \theta -9 k-2\right]+1}}.
\end{equation}
Thus, using  (\ref{eq33})  the Hubble function becomes
\begin{equation}
h=\frac{1}{6} \sqrt{\frac{\theta \left[\theta (6 (k-1) (3 k+1) \theta -9
k-1)+1\right]}{2 \theta \left[12 (k+1) (3 k+1) \theta -9 k-2\right]+1}},
\end{equation}
which leads to a scale factor of the form 
\begin{equation}
a=a_0\theta ^{-1/6}.
\end{equation}
Additionally, using (\ref{qq}) the deceleration parameter becomes
\begin{equation}
q=2+\frac{3 \left[(9 k+1) \theta -2\right]}{\theta \left[6 (k-1) (3 k+1)
\theta -9 k-1\right]+1}
+\frac{6-6 (9 k+2) \theta}{2 \theta \left[12 (k+1) (3 k+1) \theta -9 k-2%
\right]+1}.
\end{equation}

\begin{figure*}[tbp]
\centering
\includegraphics[width=8cm]{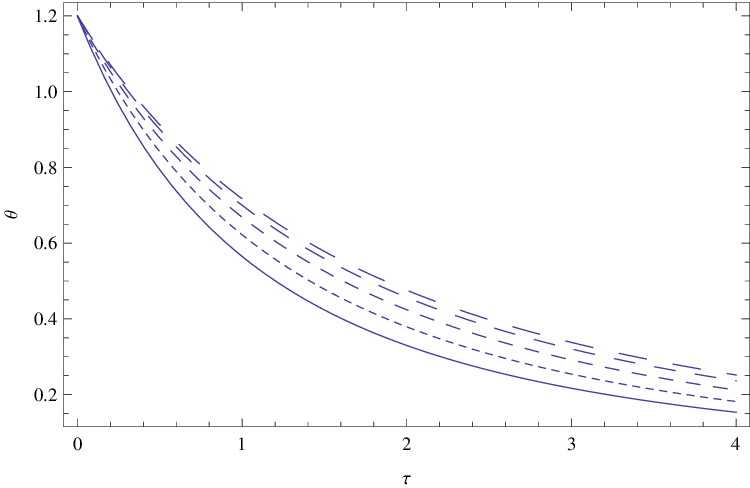}  %
\caption{{\it{Time evolution  of the matter energy density $\protect\theta $
of the stiff-fluid Universe as a function of $\protect\tau $, for different
values of $k$: $k=-1
$ (solid curve), $k=-3$ (dotted curve), $k=-5$ (short dashed curve), $k=-9$
(dashed curve) and $k=-16$ (long dashed curve). The initial value of the
energy density is $\protect\theta (0)=1.20$. }}}
\label{fig5}
\end{figure*}

\begin{figure*}[tbp]
\centering
\includegraphics[width=7.5cm]{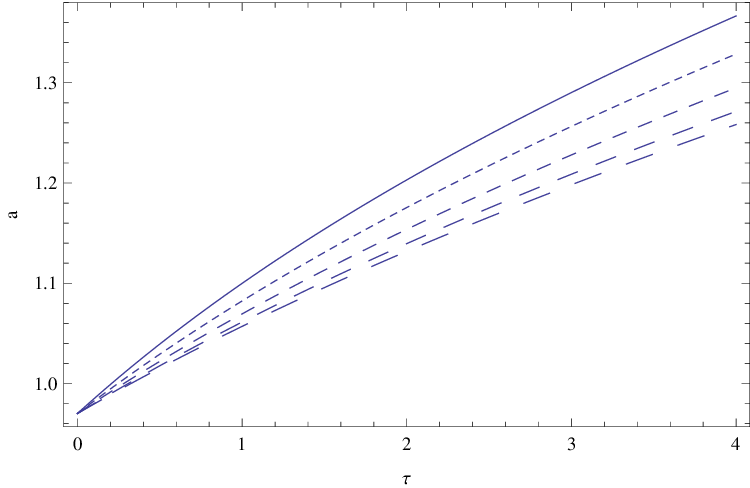}
\includegraphics[width=7.5cm]{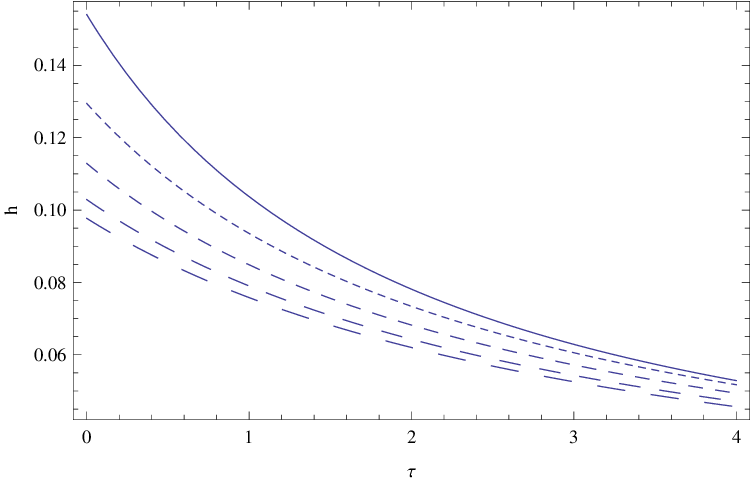}  %
\caption{{\it{Time evolution of the scale factor $a$ (left panel) and of the 
Hubble function $h$  (right panel) of the stiff-fluid
Universe as a function of $\protect\tau$, for different values of $k$:
$k=-1$
(solid curve), $k=-3$ (dotted curve), $k=-5$ (short dashed curve), $k=-9$
(dashed curve) and $k=-16$ (long dashed curve). The initial value of the
energy density is $\protect\theta (0)=1.20$. }}}
\label{fig6}
\end{figure*}

\begin{figure}[tbp]
\centering
\includegraphics[width=7.5cm]{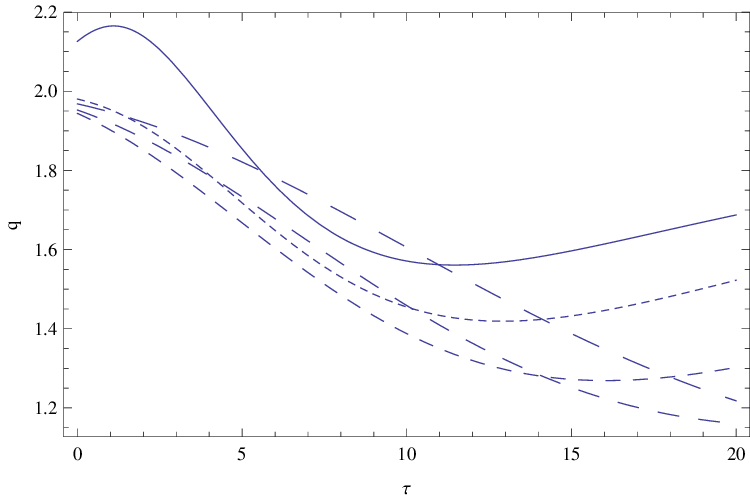}
\includegraphics[width=7.5cm]{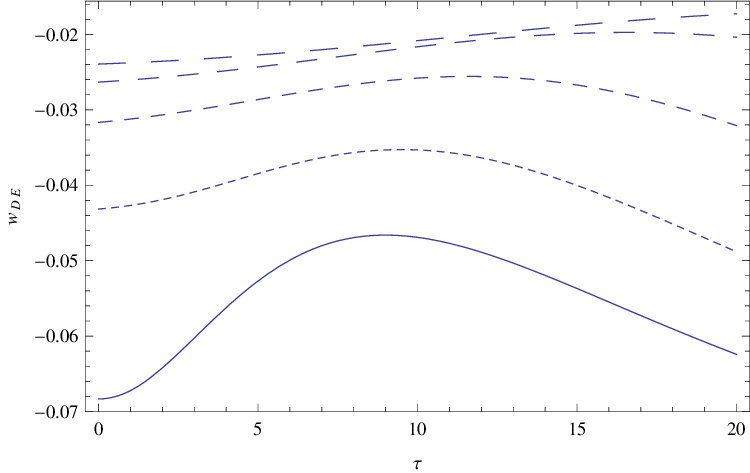}  %
\caption{{\it{
Time evolution of the deceleration parameter $q$ (left panel) and of the 
dark-energy equation of state parameter $w_{DE} $  (right panel) of the
stiff-fluid Universe as a function of $\protect\tau$, for different values of
$k$: $k=-1$ (solid curve), $k=-3$ (dotted curve), $k=-5$ (short dashed
curve), $k=-9$ (dashed curve) and $k=-16$ (long dashed curve). The initial
value of the energy density is $\protect\theta (0)=1.20$. }}}
\label{fig7}
\end{figure}

We numerically evolve the scenario at hand for various parameter values, and
in Figs.~\ref{fig5}-\ref{fig7} we respectively depict the time evolution
of  the matter energy density, of the scale
factor, of the scaled Hubble function, of the deceleration parameter and of
the  dark-energy equation-of-state parameter.

As we observe, for the considered numerical values of the model parameters 
the cosmological evolution of the Universe is strongly decelerating for all
times, with the numerical values of the deceleration parameter in the range
$1.2<q<2.2$. This is an expected result, since we have neglected the presence
of the cosmological constant $\Lambda $ in the field equations. In physical
terms it is also expected, since from the cosmological point of view the
stiff-fluid regime lasts for a very short period during the evolution of
the early Universe, in which the Zeldovich density conditions are satisfied.
During this period the Universe is expanding at a slow rate, with the Hubble
function monotonically decreasing in time. Furthermore, the dark-energy
equation-of-state parameter has small negative values, being very
close to zero for all considered dimensionless times. 

At this point, let us remind that in
standard General Relativity the evolution of the stiff-fluid Universe
is given by $\rho _{GR}=\rho _0/a^6$, $\dot{H}_{
GR}=-\rho_{GR}$, and $q_{GR}=2$. Hence, we deduce that the presence of the
higher derivative matter terms in the gravitational field equations
leads to a significant departure from the standard cosmological dynamics.
This feature could lead to strong nucleosynthesis constraints on the present
scenario, and used to distinguish it from other modified gravity classes.

\section{Conclusions}
\label{Conclusions}

In the present work we have investigated the cosmological implications of a
new class of modified gravity, which arises from the introduction of
non-dynamical auxiliary fields. This feature leads the equations of motion to
contain higher order derivatives of the matter fields. Although this could
place tight observational constraints on this theory, as long as these
constraints are satisfied the above class corresponds to a novel modified
gravitational theory and cosmology that is worthy to explore. In particular,
extracting the
gravitational field equations and imposing a flat, homogeneous and
isotropic geometry, we obtained the Friedmann equations, in which
the effective dark-energy sector contains higher derivatives of the matter
energy density and pressure.
 
One important feature of the present scenario is the conservation of the
matter energy-momentum tensor, which imposes strong constraints on the
cosmological evolution. In particular, due to this conservation  the
density - scale-factor relation is the same as in standard General
Relativity,
which in the case of a barotropic cosmological matter fluid with $p=(\gamma
-1)\rho $ takes the form $\rho \propto a^{-3\gamma }$. However, since the
evolution of the scale factor is now different, the
matter-density time evolution differs from  the standard General Relativistic
one.

In the case where the matter sector is dust-like ($p=0$), the Universe
approaches asymptotically the de Sitter stage
with $H\approx H_0={\rm constant}$ and with the matter content decreasing
according to $\rho \approx \rho _0\exp\left(-3H_0t\right)$, 
where both the deceleration parameter $q$ and the dark-energy
equation-of-state parameter $w_{DE}$ tend to $-1$, due to the domination of
the explicit cosmological constant. This result is independent of the model
parameters as expected, since in all cases the decrease of the matter
energy density due to the expansion leads all the terms in the
matter-dependent effective dark-energy sector to disappear, apart from the
cosmological constant one. Hence, in the presence of
the cosmological constant the de Sitter stage is an attractor solution of the
field equations. This was actually expected, since as we stated in subsection 
\ref{model2.1}, in the case of vanishing energy-momentum tensor the new non-GR terms 
vanish too, which is indeed the case in the above asymptotic regime. We stress however 
that at intermediate times the scenario at hand can be very different than GR. Finally, 
on the other hand, in the absence of an explicit
cosmological constant, the Universe results in a non-accelerating,
matter-dominated Universe.

In the case of radiation matter $p=\rho /3$, and in the absence of an
explicit cosmological constant we obtained a non-accelerating Universe,
similar to the radiation-dominated phase of standard General Relativity.
Additionally, in the case of a stiff matter  $p = \rho $, which is expected
to be realized in the very early Universe, we found that the Universe is
expanding at a slow rate, with the effective  dark-energy
equation-of-state parameter having small negative values.

The above variety of cosmological evolutions reveals that the theory with
higher matter derivatives is indeed new, and deserves further investigation.
More specifically, it would be interesting and necessary to perform a
detailed cosmological perturbation analysis, in order to examine the
stability properties of the theory, as well as to confront it with
perturbation-related observables such as the large-scale structure and
the growth index. Moreover, one could perform a full phase space analysis,
which will reveal the global, asymptotic behavior of the scenario.
Furthermore, another important avenue of analysis
would be to use data from Type Ia Supernovae (SNIa), Baryon
Acoustic Oscillations (BAO), and Cosmic Microwave Background
(CMB) observations, in order to impose constraints on the theory. The above
investigations may also provide specific signatures and effects, which could
distinguish between this theory and other alternatives of modified
gravity. We aim to explore in detail these issues in upcoming
publications.

\section*{Appendix}

\appendix

\section{Divergence-free character of the tensor $S_{ab}$}\label{app}

% \section*{Appendix}

In this Appendix, we will deduce the conditions imposed in order for the tensor $S_{ab}$ 
to be divergence-free, namely $\nabla_a S^{ab}=0$. Indeed, the tensor $S_{ab}$ possesses 
the following properties: i) it is imposed to vanish in vacuum, $T_{ab}=0$; ii) it is 
divergence-free, due to the Bianchi 
identity, $\nabla_a G^{ab}=0$, and the fact that $T_{ab}$ is also divergence-free, namely
$\nabla_a T^{ab}
=0$. Note that the latter property, in this theory, arises due to the fact that matter is 
minimally coupled to the metric, which guarantees $\nabla_a T^{ab}=0$. Thus, these 
theories, that possess auxiliary fields, correspond to modifying the Einstein field 
equation by adding a divergence-free tensor that vanishes in vacuum and that depends on 
the metric, the energy-momentum tensor and its derivatives.

Now, it is important to emphasize that the form of $S_{ab}$, considered in this work and 
in Ref. \cite{Pani:2013qfa}, is given as an expansion in orders of the matter fields $T$ 
and their derivatives, namely equation (\ref{eqexp}) is given in terms of a derivative 
expansion. Hence, it is by definition approximate and not exact. One cannot expect to get 
exact vanishing of the divergence precisely for this reason. Instead, what is important is 
that order by order in the expansion $\nabla_a S^{ab}$ is zero to that order. For 
instance, one maintains terms which are $O(T^2)$, while terms such as $\Lambda \beta_1^2 
\nabla^b (T^2)$ contribute to higher orders. Thus, if one assigns two derivatives to $T$ 
then one essentially keeps up to four derivatives. More specifically, consider $k$ a 
parameter that counts derivatives in this way, then indeed one has $\nabla_a S^{ab}
=0+ O(k^6)$ which is all that is needed, as $O(k^6)$ terms were not included anyway in the 
original expansion (note that this is a standard procedure in any perturbation theory).

In this context, the tensor $S_{ab}$, up to fourth order in the derivatives, is thus given 
by equation (\ref{eqexp}), which we reproduce here
\begin{eqnarray}
S_{ab} &=& \alpha_1\, g_{ab}\,T  
 + \alpha_2\, g_{ab}\, T^2 + \alpha_3 \, T\, T_{ab} + \alpha_4\, g_{ab}\, T_{cd}\,T^{cd}  
 + \alpha_5 \, T^c\,_a\,T_{cb}  
    \nonumber  \\
 &&+ \beta_1\, \nabla_a\nabla_b\,T + \beta_2 \, g_{ab}\, \Box\,T   +  \beta_3\, 
\Box\,T_{ab} + 2\beta_4\, \nabla^c\nabla_{(a}\,T_{b)c}+\ldots\,.
 \textsc{\label{eqexpapp}}  
\end{eqnarray}
In order to impose that $S_{ab}$ is divergence-free, to the required order, the following 
relations are imposed:
\begin{eqnarray}
(\Box\,\nabla_b-\nabla_b\,\Box)\,T &=& R_{ab} \nabla^a\,T,  \label{rel2a} \\
(\nabla^a\,\nabla^c\,\nabla_a-\nabla^c\,\Box)T_{cb} &=& R_{abcd}\nabla^d\,T^{ca},  
\label{rel2b}\\
\nabla^a R_{abcd} &=& 2\nabla_{[c}R_{d]b}\,, \label{rel2c}
\end{eqnarray}
and taking into account the lowest-order equation given by $R_{ab}=T_{ab}-\alpha 
g_{ab}T+g_{ab}\Lambda + ...$ (once again, we refer the reader to Ref. \cite{Pani:2013qfa} 
for more details), lead to the necessary cancellations between terms to $O(k^6)$. Note, 
however, that equations (\ref{rel2a})-(\ref{rel2c}) are exact identities.

Finally, the resulting modified Einstein equation takes the form of equation 
(\ref{fieldeq}), which 
we reproduce here
\begin{eqnarray}
G_{ab} &=& T_{ab} - \Lambda g_{ab}      -
\beta_1 \Lambda\, g_{ab}\,T + \frac{1}{4}\left(1-2\beta_1
\Lambda\right)(\beta_1-\beta_4)\, g_{ab}\, T^2  \notag \\
&&
+ \left[\beta_4\left(1-2\beta_1\, \Lambda\right) - \beta_1\right] \, T\,
T_{ab} + \frac{1}{2}\,\beta_4 \, g_{ab}\, T_{cd}\, T^{cd}  - 2\beta_4\,
T^c\,_a\,T_{cb} \notag \\
&& + \beta_1\, \nabla_a\nabla_b\,T - \beta_1\,
g_{ab}\, \Box\,T  - \beta_4 \, \Box\,T_{ab} + 2\beta_4\,
\nabla^c\nabla_{(a}\,T_{b)c}
+\ldots\,.  \label{fieldeqapp}
\end{eqnarray}

\begin{acknowledgments}
The authors are grateful to Paolo Pani and Thomas Sotiriou for helpful clarifications and 
comments.FSNL acknowledges financial  support of the Funda\c{c}\~{a}o para a
Ci\^{e}ncia e Tecnologia through an Investigador FCT Research contract, with
reference IF/00859/2012, funded by FCT/MCTES (Portugal), and grants
CERN/FP/123618/2011 and EXPL/FIS-AST/1608/2013. The research of ENS is
implemented within the framework of the Operational Program ``Education and
Lifelong Learning'' (Actions Beneficiary: General Secretariat for Research
and Technology), and is co-financedby the European Social Fund (ESF) and the
Greek State.
\end{acknowledgments}

\end{document}